\documentclass[twocolumn,pra,aps,showpacs]{revtex4}
\usepackage{xspace,amsmath,amsfonts,amsthm,amssymb,amsbsy,graphicx,color}

\begin{document}

\newcommand{\smallfrac}[2]{\mbox{$\frac{#1}{#2}$}} 
\newcommand{\half}{\smallfrac{1}{2}}
\newtheorem{lemma}{Lemma}

\title{Feedback control for communication with non-orthogonal states}

\author{Kurt Jacobs}

\affiliation{Quantum Science and Technologies Group, Hearn Institute for Theoretical Physics, Department of Physics and Astronomy, 202 Nicholson Hall, Tower Drive, Baton Rouge, LA 70803, USA}

\begin{abstract}

Communicating classical information with a quantum system involves the receiver making a 
measurement on the system so as to distinguish as well as possible the alphabet of states 
used by the sender.  We consider the situation in which this measurement takes an 
appreciable time. In this case the measurement must be  described by a continuous 
measurement process. We consider a continuous implementation of the optimal measurement 
for distinguishing between two non-orthogonal states, and show that feedback control 
can be used during this measurement to increase the rate at which the information 
regarding the initial preparation is obtained. We show that while the maximum 
obtainable increase is modest, the effect is purely quantum mechanical in the sense 
that the enhancement is only possible when the initial states are non-orthogonal. We 
find further that the enhancement in the rate of information gain is achieved at the 
expense of reducing the total information which the measurement can extract in the 
long-time limit.

\end{abstract}


\maketitle

\section{Introduction}

A quantum channel is a quantum system which is sent from one person (the sender) to another (the receiver).  The sender prepares the system in one of a set (or alphabet) of states, and the receiver makes a measurement upon the system to determine as best as possible which state the sender has prepared. For practical purposes one is interested in the maximum rate at which the sender and receiver can use this protocol to transmit information reliably in the limit in which a sequence of many quantum systems are sent. The better the receiver can distinguish between the senders states, the more information can be sent per system (per use of the channel). When the receiver makes separate measurements on each system in the sequence (which is the situation we consider here), the amount of information that can be sent per use of the channel is given by the {\em mutual information} which is generated between the sender and receiver by the receiver's measurement~\cite{KR,H96,H,SW}. 

The problem of communication is closely related to that of state-discrimination~\cite{C00}; the mutual information is a measure of how well the receiver is able to distinguish the states in the senders set. When two or more of the states in the set are non-orthogonal, it is not possible to do this perfectly - at best there will always be a non-zero probability that the observer will remain to some degree unsure about the initial preparation. The main difference is that in communication the objective is to maximize the mutual information, whereas in state-discrimination other measures of success are also considered. An example is the problem of {\em unambiguous discrimination} between a set of states~\cite{I87}. In this case the objective is to maximize the probability that the observer obtains full information regarding the initial preparation. 

Here we will be concerned with discriminating between two non-orthogonal pure states, and will be exclusively concerned with the mutual information. Specifically the problem is as follows: The receiver knows that one of the two states $\{|\psi_1\rangle,|\psi_1\rangle\}$ has been prepared, with the respective probabilities $\{P_1,P_2\}$.  She wishes to make a measurement so that her final state of knowledge regarding which state was prepared, $\{P_1',P_2'\}$ has as low a Shannon entropy as possible. That is, she wishes to minimize $H(\{P_i'\}) = -\sum_i P_i'\ln(P_i')$. This problem was first solved by Levitin~\cite{Levitin} (see also~\cite{FuchsCaves,FuchsPhD}), and the solution is quite intuitive. Since only two pure states are involved, we need only consider a two-dimensional state space, and thus the Bloch sphere is sufficient for describing the problem. Without loss of generality, we may assume that the two states lie in the Bloch sphere's $x$-$z$ plane, and that they are placed symmetrically on either side of the $z$-axis. In this case the optimal measurement is one which projects the system onto either one of the two $\sigma_x$ eigenstates. (To state this in a more general fashion, the optimal measurement is one which projects the system onto a basis that is perpendicular to the direction which bisects the angle between the coding states.) The measurement in question is a von Neumann measurement described by the two projectors $\{P_\pm = |\pm\rangle_x\langle\pm|_x\}$, where $|\pm\rangle_x$ are the two eigenstates of $\sigma_x$. To obtain the maximum mutual information between the sender and receiver, the sender must choose the initial probabilities to be $P_1 = P_2 = 1/2$. If we write the two coding states as 
\begin{eqnarray}
  \rho_1 = \frac{1}{2}[I + \sin(\theta)\sigma_x + \cos(\theta)\sigma_z ] , \label{code1} \\ 
  \rho_2 = \frac{1}{2}[I - \sin(\theta)\sigma_x + \cos(\theta)\sigma_z ] ,
  \label{code2}
\end{eqnarray}
then $\theta$ is the angle that the Bloch vector of each state makes with the $z$-axis, and the mutual information generated between the sender and receiver by the optimal measurement is 
\begin{eqnarray}
 M_{\mbox{\scriptsize opt}} = \ln2 +  \sum_{\pm}\left[\frac{1}{2} \pm \frac{1}{2} \sin(\theta)\right] \ln\left[\frac{1}{2} \pm \frac{1}{2}\sin(\theta)\right] .
\end{eqnarray}

While measurements are often treated as being instantaneous for simplicity, all measurements take a finite time. If this time is not short compared to other important time scales in a given problem, then it is important to determine the time-dependence of the information provided by the measurement, and this requires examining the continuous dynamics of the measurement process. When sending information over a quantum channel, if the receiver's measurement time is not small compared to the senders preparation time, then the measurement time will affect the rate of transmission (the capacity) of the channel. Here we consider such a situation, and show that feedback control can be used by the receiver during the measurement to increase the transmission rate. 

To understand why this is the case, it is helpful to approximate the continuous measurement by a sequence of weak measurements, each of which extracts a little information regarding which of the eigenstates of $\sigma_x$ the system is in. Such a weak measurement, which we will denote by ${\cal M}(k)$, is described by the two operators
\begin{equation}
   \Omega_\pm = \frac{1}{2}(\sqrt{k}+\sqrt{1-k}) I \pm ( \frac{1}{2}(\sqrt{k}-\sqrt{1-k}) \sigma_x .
\end{equation}
When $k$ is close to $1/2$, the measurement operators $\Omega_\pm$ are close to the identity, and almost no information is extracted regarding the state of the system. On the other hand, when $k=0$ or $1$, the measurement projects the system onto one of the eigenstates of $\sigma_x$, and is therefore the optimal measurement described above. If we choose $k\in (0,1)$, then the optimal measurement is obtained in the limit when one  repeats this weak measurement many times.  A continuous measurement is obtained by setting $k = 1/2 - \sqrt{\varepsilon\Delta t}$, for some fixed $\varepsilon$, making the resulting weak measurement in each time interval $\Delta t$, and taking the continuum limit $\Delta t \rightarrow dt$~\cite{B02}. 

Now consider what happens when we make a sequence of these weak measurements ${\cal M}(k)$. For concreteness we will set $k$ to be in the interval $(0.5,1)$. Let us assume that the result of the first measurement is that associated with the operator $\Omega_+$. In this case, this operator is applied to each of the coding states, so that after the measurement these states are transformed to $\rho_i' = \Omega_+ \rho_i \Omega_+^\dagger/\mbox{Tr}[\Omega_+^\dagger \Omega_+ \rho_i]$. The effect of this is to rotate both of the coding states towards the $\sigma_x$ eigenstate $|+\rangle_x$. Because of the asymmetry of the states with respect to $|+\rangle_x$, the two states are not rotated by the same amount; after the measurement the coding states are closer together than they were initially. This reflects the fact that one cannot extract unlimited information by repeating the measurement many times -- each time the measurement is repeated, the coding states are brought closer together, so that each subsequent measurement extracts (on average) less information than its predecessor.  

Because both the coding states are rotated towards $|+\rangle_x$, they are no longer symmetrically placed about the $z$-axis. Now, as we described above, the optimal measurement for extracting information from two non-orthogonal states is one that projects the system onto states in a basis which is perpendicular to the direction which bisects the angle between the non-orthognal states. Similarly, our weak measurement will extract the most information when its operators are diagonal in a basis which is perpendicular to the direction bisecting the coding states. However, because of the rotation resulting from the first measurement, this basis is no longer the $\sigma_x$ basis, and as a result the second measurement is not oriented so as to obtain the maximum information. To do so we therefore need to rotate the system (or the measurement), so that the symmetry is restored. Since the direction of the required rotation depends upon the outcome of the previous measurement, this procedure consists of a process of {\em feedback control}~\cite{B88,Bb88,WM93b,YK98,DJ99,DHJMT00,Jacobs03,SGDM,MJ04,B05,CJ06,DJxx}. 

It is important to note that in the absence of this feedback procedure, even though each subsequent measurement is not oriented so as to extract the maximum possible information, this does not prevent the sequence of measurements from extracting the optimal information in the limit of many repetitions. They must do so, since as described above they produce the optimal von Neumann measurement in this limit. However, each measurement in the sequence individually does not extract as much information as it could do. The feedback procedure thus increases the {\em rate} at which the information is obtained by the sequence, not the total final information that the sequence extracts. In fact, interestingly and unexpectedly, we will find that while increasing the initial rate of information gain, the feedback algorithm actually {\em decreases} the total information which is extracted in the long-time limit. That is, the rotation caused by each of the individually optimal weak measurements is such as to close the gap between the coding states more (on average) than is strictly necessary as a result of the information that each, on average, obtains.

In the following section we solve the stochastic equations which govern the evolution of the receivers state-of-knowledge regarding which of the coding states has been prepared, and calculate the time evolution of the resulting mutual information. In Section~\ref{fbsec} we determine the feedback algorithm which maximizes the information gain in each infinitesimal time-step (given the current state at each time), and solve for the evolution of the receivers state of knowledge under this feedback.  We compare the resulting mutual information to that without feedback, and calculate the increase in the transmission rate which the algorithm provides. Section~\ref{conc} concludes with a brief summary of the results.

\section{The dynamics of continuous state-discrimination}
A simple continuous version of a von Neumann measurement that projects a two-state system onto one of the eigenvalues of $\sigma_x$, and one that is widely applicable, is described by the stochastic differential equation~\cite{Jacobs03}
\begin{equation}
   d\rho = -\gamma [\sigma_x,[\sigma_x,\rho]] dt + \sqrt{2\gamma} ( \sigma_x \rho + \rho \sigma_x - 2\langle\sigma_x\rangle \rho ) dW ,
   \label{sme}
\end{equation} 
where $\rho$ is the state of the quantum system evolving under the measurement, and $dW$ is an infinitesimal increment of Gaussian white noise, referred to as the {\em Wiener process}~\cite{Gilespie}. The continuous stream of output results, which we will denote by $y(t)$, is determined by $dy =  \langle \sigma_x \rangle dt + dW/\sqrt{8\gamma}$. The parameter $\gamma$ determines the rate at which the measurement extracts information, and is referred to as the measurement rate or the {\em strength} of the measurement. Examples of explicit implementations of this measurement on solid-state qubits are given in~\cite{K99}.

In our case at the start of the measurement the system has been prepared in a mixture of the coding states, so that $\rho(0) = \sum_i P_i \rho_i$, with the $\rho_i$ given by Eqs.(\ref{code1}) and (\ref{code2}). We wish to calculate both the evolution of the receiver's state of knowledge regarding which of the coding states was prepared (it is this that tells us the mutual information generated between sender and receiver), and the dynamics of each of the coding states which make up the mixture. 

Given Eq.(\ref{sme}), we can immediately write down the stochastic equations for the coding states: If the receiver was making a measurement on a system in coding state $\rho_i$, then the equation of motion for $\rho_i$ would simply be given by Eq.(\ref{sme}) with $\rho$ replaced with $\rho_i$. However, since the receiver's state is actually $\rho$, and the receiver is making a single measurement which evolves both $\rho$ and the two $\rho_i$, we must use the record generated by the equation for $\rho$ when we evolve the equations for the $\rho_i$. Substituting this record into the equations for the $\rho_i$, we obtain
\begin{eqnarray}
   d\rho_i   & = &   -\gamma [\sigma_x,[\sigma_x,\rho_i]] dt \nonumber \\ 
                 &    &  + \sqrt{2\gamma} ( \sigma_x \rho + \rho \sigma_x - 2\mbox{Tr}[\sigma_x\rho_i] \rho_i ) dW_i , \\
   dW_i      & = &  dW - \sqrt{8\gamma}( \mbox{Tr}[\sigma_x\rho] - \mbox{Tr}[\sigma_x\rho_i]) dt 
   \label{smecodes}
\end{eqnarray}

To derive the equation of motion for the receiver's state-of-knowledge regarding the initial preparation, we can use Bayes' theorem~\cite{Bayes}. To do this the quantity we require is the conditional probability for the measurement outcome (record) given the choice of coding state. We can obtain this directly from the measurement records associated with the equations for the $\rho_i$, and this gives
\begin{eqnarray}
   P(dy|i) & = & \sqrt{\frac{4\gamma}{\pi}} e^{-4\gamma( dy - \mbox{\scriptsize Tr}[\sigma_x \rho_i]dt )^2} \nonumber \\
              & = & 1 + \mbox{Tr}[\sigma_x\rho_i] ( 8\gamma \mbox{Tr}[\sigma_x\rho] dt + \sqrt{8\gamma} dW) 
\end{eqnarray}
Note that we must be careful to expand everything to second order in $dW$ because $dW^2=dt$~\cite{Gilespie}. Using Bayes' theorem, the update to the probabilities $P_i$, given the measurement record is 
 \begin{eqnarray}
   P(i|dy) & = & \frac{P(dy|i)P_i}{\sum_i P(dy|i)P_i} \\
              & = & P_i ( 1 + \sqrt{8\gamma}\{\mbox{Tr}[\sigma_x\rho_i] - \mbox{Tr}[\sigma_x\rho]\}dW) .
\end{eqnarray}
The equation of motion for the $P_i$, which describe the receiver's state of knowledge regarding the initial preparation, is therefore 
\begin{equation}
   dP_i = \sqrt{8\gamma}(\mbox{Tr}[\sigma_x\rho_i] - \mbox{Tr}[\sigma_x\rho])dW P_i .
   \label{Ps}
\end{equation}
Equations (\ref{sme}), (\ref{smecodes}) and (\ref{Ps}) are a complete description of the state-discrimination process under the continuous measurement. Note that at all times $\rho(t) = \sum_i P_i(t)\rho_i(t)$, so one can check the equations for the $d\rho_i$ and $dP_i$ by verifying that they generate the correct equation for $d\rho$. 

\begin{figure}
\leavevmode\includegraphics[width=0.95\hsize]{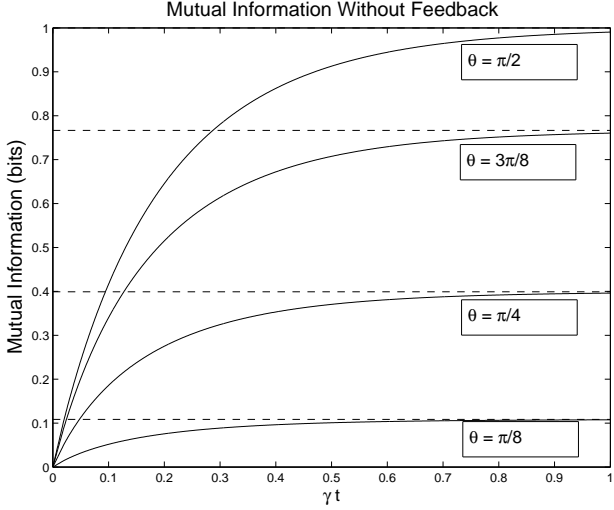}
\caption{The solid lines give the mutual information as a function of the measurement time, for a continuous measurement in the optimal basis, and for various values of $\theta$, being half the angle separating the coding states. The dashed lines give the maximum information which can be extracted for each value of $\theta$. } \label{fig1}
\end{figure}

While these equations are non-linear, and at first sight may look intractable, we can obtain a closed form solution using the method which involves transforming the equations to their equivalent linear form~\cite{GG}. Note that since the equation for $\rho$ is stochastic, the solution consists of a set of possible final states, with a probability for each one. The solution for $\rho$ is (see, e.g.~\cite{JK97,CJ06})
\begin{equation}
    \rho(t) = \frac{e^{4\gamma t\sigma_z v} \rho(0) e^{4\gamma t\sigma_z v}}
                         {\mbox{Tr}[e^{8\gamma t\sigma_z v} \rho(0)]} ,
\end{equation}
where $v$ parameterizes the possible states at time $t$, and the probability density for $v$ at time $t$ is
\begin{equation}
    P(v,t) = \sqrt{\frac{4\gamma t}{\pi}} e^{-4\gamma t (v^2 + 1)} \mbox{Tr}[e^{8\gamma t\sigma_z v} \rho(0)] .
\end{equation}
Substituting $\rho(t)=\sum_i P_i(t)\rho_i(t)$ into this equation gives the solution for the $P_i$ which is
\begin{equation}
   P_i(t) = \frac{\mbox{Tr}[e^{8\gamma t\sigma_z v} \rho_i(0)] P_i(0)}
                      {\mbox{Tr}[e^{8\gamma t\sigma_z v} \rho(0)]} .
\end{equation}
We are interested in the mutual information at time $t$, which is obtained by taking the diference between the entropy of the observer's initial state of knowledge, and that of her state-of-knowledge at time $t$, and averaging this over all the possible outcomes at that time. This is 
\begin{equation}
   M(t) = H(\{P_i(0)\}) - \int_{-\infty}^{\infty} H(\{P_i(t)\}) P(v,t) dt . 
   \label{mi}
\end{equation}
Choosing the initial probabilities to be $P_1=P_2=1/2$, we evaluate this integral numerically, and plot the result for various values of $\theta$ in Figure~\ref{fig1}. We see from this that the information obtained by the continuous measurement as $t\rightarrow\infty$ tends to the maximal value as expected. 

\section{Optimal feedback control} \label{fbsec}
We now wish to apply feedback to the system during the measurement so that in each infinitesimal time interval the measurement obtains the maximal amount of information possible given the state of the system at that time. As discussed in the introduction, this can be done by performing a rotation of the system in the $x$-$z$ plane in each infinitesimal interval so that the coding states remain symmetrically placed about the $z$-axis. 

To calculate the required angle of rotation we first calculate the equations of motion for the $x$ and $z$ components of the Bloch vector of the coding states, which are
\begin{eqnarray}
   dz_i & = &  (+\sqrt{8\gamma} z_i dW) x_i   \nonumber \\ 
           &   & - 8\gamma z_i dt  \left( 1/2 - x_i \left\{ x_i - \sum_j [P_j x_j] \right\}  \right) , \\ 
   dx_i & = &  (-\sqrt{8\gamma} z_i dW) z_i   \nonumber \\ 
           &    &  - 8\gamma z_i^2 dt  \left( x_i - \sum_j [P_j x_j] \right)    .
\end{eqnarray}
Before the measurement in each infinitesimal step the coding states are symmetrically placed about the $z$-axis, so that $z_1=z_2=z$ and $x_1=-x_2=x$. After the measurement we need to effect an infinitesimal rotation so as to restore these conditions. First let us denote the increments in $z$ and $x$ due to a rotation by an angle $\phi$ as $dz^{\mbox{\scriptsize R}}(\phi,x,z)$ and $dx^{\mbox{\scriptsize R}}(\phi,x,z)$, respectively. Second we denote the values of $x_i$ and $z_i$ after the measurement in the interval $dt$ by $x_i' = x_i + dx_i$ and $z_i' = z_i + dz_i$. We now impose the condition that the sum of the increments due to the measurement and rotation restore the relations $z_1 = z_2$ and $x_1 = -x_2$. The resulting conditions which the rotation must satisfy are
\begin{eqnarray}
  dz^{\mbox{\scriptsize R}}(\phi,x_1',z_1') - dz^{\mbox{\scriptsize R}}(\phi,x_2',z_2') & = &  
  dz_1 - dz_2 , \\ 
  dx^{\mbox{\scriptsize R}}(\phi,x_1',z_1') + dx^{\mbox{\scriptsize R}}(\phi,x_2',z_2') & = &  
  dx_1 + dx_2 .
\end{eqnarray}
Solving these equations for $\phi$, making sure that we include all terms to second order in $dW$, gives
\begin{equation}
  \phi = \sqrt{8\gamma} dW + 8\gamma z x (P_1 - P_2) dt .
\end{equation}
The total increments in $x_1 \equiv x$ and $z_1 \equiv z = \sqrt{1-x^2}$ resulting from the measurement and subsequent rotation in each interval $dt$ are 
\begin{eqnarray}
  dz =  4\gamma z x^2 dt  , \\ 
  dx =  -4\gamma x z^2 dt  . 
\end{eqnarray}
Note that the feedback has cancelled the stochastic increments, resulting in a deterministic evolution for the coding sates. The equation of motion for the $z$ component of the Bloch vector of the coding states under the feedback algorithm is thus
\begin{eqnarray}
  \dot{z} =  4\gamma z (1-z^2)  .  
  \label{zeq}
\end{eqnarray}
Defining $\xi = \tan(\theta(t)) = x/z = \sqrt{1-z^2}/z$, the solution to this equation is 
\begin{eqnarray}
 \xi(t) = \xi(0) e^{-\gamma t} = \tan(\theta_0)e^{-4\gamma t} ,
\end{eqnarray}
where we have defined $\theta_0 = \theta(0)$.

To implement the feedback algorithm the receiver applies a time dependent Hamiltonian to the system so as to perform the rotation $\phi(t)$ in each time step $dt$. The required feedback Hamiltonian is
 \begin{equation}
  H_{\mbox{\scriptsize fb}} = 8\gamma\hbar \sigma_z  \{ y(t) +  [z(t) -1] \langle \sigma_x (t) \rangle  \}
\end{equation}
where $y(t)$ is the measurement record, and $z(t)$ is the $z$-component of the coding states at time $t$, who's deterministic evolution is given by Eq.(\ref{zeq}) above. This is a complete specification of the feedback algorithm. To implement it the receiver simply determines the feedback Hamiltonian directly from the measurement record as the measurement proceeds and applies it continually to the system. 

\begin{figure}
\leavevmode\includegraphics[width=0.95\hsize]{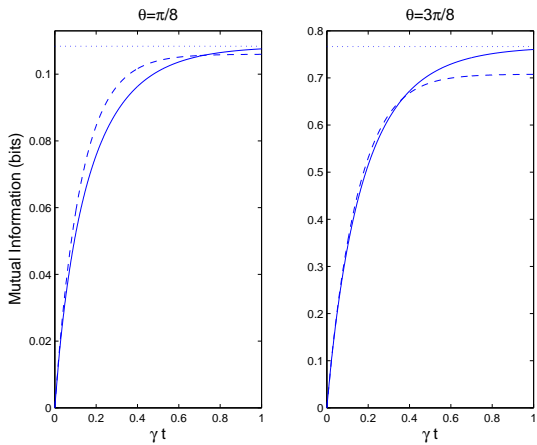}
\caption{Here we plot the mutual information as a function of the measurement time for two values of $\theta$, being half the angle between the coding states. Solid lines: measurement alone; dashed lines: measurement with feedback. The dotted lines give the maximum information which can be extracted for each value of $\theta$.} \label{fig2}
\end{figure}

To calculate the performance of the feedback algorithm, we need to solve the equation of motion for the $P_i$ under the feedback. Given the above evolution for the coding states (Eq.(\ref{zeq})), this equation becomes
\begin{eqnarray}
   dP_\pm & = & \sqrt{8\gamma}(\mbox{Tr}[x_i] - \sum_j [P_j x_j]) P_i dW   ,  \\
                 & = & \sqrt{8\gamma}\sin(\theta(t))[P_2 - P_1 \pm 1] P_i dW .
\end{eqnarray}
where for notational convenience we have defined $P_+ \equiv P_1$ and $P_- \equiv P_2$.
This is merely a classical continuous measurement on a classical two-state system, with 
the time dependent measurement strength $\gamma_{\mbox{\scriptsize c}}(t) = \gamma\sin^2(\theta(t))$. 
Apart from the fact that the measurement strength is now time dependent,  this is a special case of the stochastic master equation Eq.(\ref{sme}) in which $\rho$ is diagonal 
in the $\sigma_x$ basis. This can be solved using the method in~\cite{JK97}, and is 
\begin{equation}
   P_\pm(t) = P_\pm(0) \frac {e^{\pm 8 \Gamma(t) u}} {P_+(0) e^{ 8 \Gamma(t) u}  + P_-(0) e^{- 8 \Gamma(t) u} }
\end{equation}
where the probability density for $u$ at time $t$ is 
\begin{equation}
   P(u,t) = \sqrt{\frac{4\Gamma(t)}{\pi}} \sum_\pm P_\pm(0) e^{-4\Gamma(t)(u\pm 1)^2} ,
\end{equation}
and $\Gamma(t)$ is 
\begin{equation}
   \Gamma(t) = \int_0^t \gamma_{\mbox{\scriptsize c}}(t') dt' = \frac{1}{8} \ln \left( \frac{1 + \tan^2(\theta_0)}
                        {1+\tan^2(\theta_0)e^{-8\gamma t}} \right) .
\end{equation}
Now that we have the solution for the observer's state-of-knowledge we can use the expression in Eq.(\ref{mi}) to calculate the mutual information as a function of time. We plot this, along with the mutual information without feedback, in Figure~\ref{fig2} for $P_1=P_2=1/2$, and for two example values of $\theta_0$. We see that when the coding states are non-orthogonal the feedback algorithm causes the mutual information to rise more rapidly. The extent of the increase depends on the angle between the coding states, increasing as the states are brought closer together. From Figure~\ref{fig2} we see that the effect is appreciable when $\theta=\pi/8$, and conversely is negligible when $\theta=3\pi/8$.   Curiously, while rising more rapidly in the medium term, the mutual information eventually saturates at a value a little below that obtained by the optimal von Neumann measurement. By optimising the amount of information obtained at each successive time-step, we reduce the total information which can be obtained in the long time limit. Thus the feedback algorithm causes the angle between the codings states to close even faster than that strictly necessary as a result of the increased information gain. 

\begin{figure}
\leavevmode\includegraphics[width=0.95\hsize]{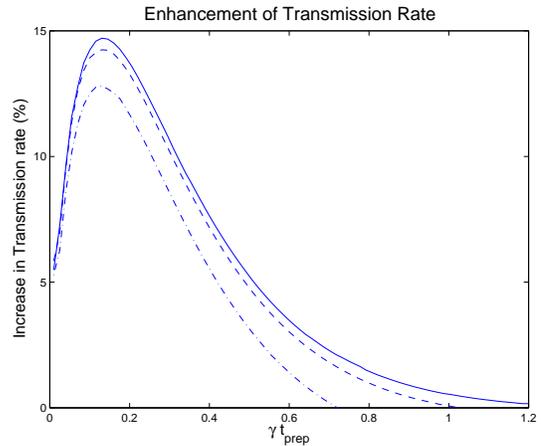}
\caption{Here we plot the percentage increase in the transmission rate of the channel due to the the feedback algorithm, as a function of the time that it takes the sender to prepare the coding  states. Solid line: $\theta=\pi/32$; dashed line: $\theta=\pi/16$; dash-dot line: $\theta=\pi/8$.} \label{fig3}
\end{figure}

We now calculate the increase in the transmission rate that can achieved by the feedback algorithm. Consider a channel in which the sender sends a sequence of two-state systems to the receiver, each one prepared in one of the two non-orthognal states $\rho_i$. The sender takes a time $t_{\mbox{\scriptsize prep}}$ to prepare each one, and the receiver has the ability to make a continuous measurement on each in turn with strength $\gamma$. The rate at which the channel can be used to send reliable information is the mutual information generated between sender and receiver by the measurement on a single system, divided by the total time taken for preparation and measurement. Since the sender can prepare the next system while the receiver is measuring the current system, the total time taken is either the preparation time or the measurement time, whichever is larger. 

The measurement time which maximizes the transmission rate depends upon the preparation time, and so we calculate this for a range of preparation times. From this is it simple to calculate the percentage increase in the optimal transmission rate provided by the feedback algorithm as a function of the preparation time. We plot this percentage increase for $\theta=\pi/8$, $\pi/16$ and $\pi/32$ in Figure~\ref{fig3} for a range of preparation times. We see that as $\theta$ is decreased, the enhancement provided by the feedback algorithm increases as expected, and saturates at about $15\%$ for small values of $\theta$.  
 
\section{Conclusion}
\label{conc}

We have considered a quantum channel in which the time taken to prepare the system and that taken to measure the system are appreciable. We have treated the case in which the receiver makes separate measurements on each system she receives, and shown that if the coding states are non-orthogonal, feedback control can used by the receiver during her measurement to increase the transmission rate of the channel. For a fixed measurement strength, $\gamma$, the amount of the increase depends upon the degree to which the coding states are non-orthogonal, and the senders preparation time. The maximum increase which can be obtained is approximately $15\%$, and this is realised when the angle between the coding states is small, and the preparation time is approximately $0.13/\gamma$.  Appreciable increases can be obtained for a range of values of $\theta$. As an example, a maximum increase of $12\%$ or more is obtained for values of $\theta$ less than or equal to $\pi/8$.  

It will be interesting to ask whether this effect remains when measurements are made on multiple systems, and whether or not the effect is larger when more than two coding states are used for communication with higher dimensional systems. 

\section*{Acknowledgments}
 This work was supported by The Hearne Institute, The National Security Agency, The Army Research Office and The Disruptive Technologies Office.


\begin{thebibliography}{99}
\bibitem{measnote}
  An good introduction to continuous quantum measurement is given in~\cite{B02}.
  Other recent treatments of continuous measurements of Hermitian operators are
  given in~\cite{DJ99,K00}, where the first uses the results of optical continuous
  measurements in~\cite{C93,WM93}. The stochastic master equation describing a 
  continuous measurement was first derived by Belavkin in~\cite{B87}, and a selection 
  of the early work on continuous measurement is given in references~\cite{BS88,D86,Bar87}.
\end{thebibliography}
\end{document}